\documentclass[12pt]{article}
\usepackage{graphicx} 
\usepackage{amsmath,amssymb,amsthm}
\usepackage{fullpage}

\usepackage{todonotes}

\usepackage{hyperref}

\newcommand{\bE}{\mathbb{E}}
\newcommand{\R}{\mathbb{R}}
\newcommand{\bI}{\mathbb{I}}

\newcommand{\tip}{\mathrm{tip}}
\newcommand{\EWS}{\mathrm{EWS}}

\DeclareMathOperator{\bump}{bump}

\title{Early warning skill, extrapolation and tipping for accelerating cascades}

\author{Peter Ashwin\thanks{Department of Mathematics and Statistics, University of Exeter, Exeter EX4 4QF, UK} \and Robbin Bastiaansen\thanks{Institute for Marine and Atmospheric research Utrecht, Department of Physics, Utrecht University, Netherlands} \and Anna S. von der Heydt${}^{\dag}$ \and Paul Ritchie${}^{*}$}
\date{January 2025}

\begin{document}

\maketitle

\begin{abstract}
We investigate how nonlinear behaviour (both of forcing in time and of the system itself) can affect the skill of early warning signals to predict tipping in (directionally) coupled bistable systems when using measures based on critical slowing down due to the breakdown of extrapolation. We quantify the skill of early warnings with a time horizon using a receiver-operator methodology for ensembles where noise realisations and parameters are varied to explore the role of extrapolation and how it can break down. We highlight cases where this can occur in an accelerating cascade of tipping elements, where very slow forcing of a slowly evolving ``upstream'' system forces a more rapidly evolving ``downstream'' system. If the upstream system crosses a tipping point, this can shorten the timescale of valid extrapolation. In particular, ``downstream-within-upstream'' tipping will typically have warnings only on a timescale comparable to the duration of the upstream tipping process, rather than the timescale of the original forcing.
\end{abstract}

\tableofcontents

\section{Introduction}

Nonlinear systems subjected to a slow varying input parameter (external forcing) can respond to forcing with abrupt behaviours even for arbitrarily slow forcing. This includes periods of rapid change due to stable states ceasing to exist (B-tipping), failure to track an attractor as a parameter is changed too rapidly (R-tipping) or due to large fluctuations in stochastic forcing (N-tipping) \cite{Ashwin2012tipTypes}. We focus here on tipping points or critical transitions that arise due to B-tipping, i.e. for slow variation of a forcing in the system and only low amplitude noise. The mathematics of tipping points in nonautonomous systems has applications in climate and ecological systems, where nonlinear feedbacks can result in periods of rapid change in response to slow natural or anthropogenic changes \cite{Scheffer2009EWSNature,Lenton2012CSDClimate}.

Driven by the need to understand the predictability of and risks associated with tipping points \cite{Lenton2008,armstrong2022exceeding}, there has been particular interest in finding {\em early warning signals} (EWS) or {\em precursors} of impending tipping points. {\em Critical slowing down} (CSD) in response to stochastic variability is one tool that can warn of an approach to such a tipping point \cite{Dakos2018MultDResiliencevsEigenvalues_EOF,George2023EWS}. Early warning signals using CSD consider an observation that is examined for signs of loss of stability that may be associated with approach to a bifurcation. This loss of stability is signalled by an increase in the variance and an approach of the lag-one autocorrelation towards $1$ \cite{Dakos2008CSDClimate,Scheffer2009EWSNature}; see Section~\ref{sec:estimation} for more details. We take the approach that the return time is a more fundamental measure of stability than the autocorrelation, as this is independent of sampling rate. We refer the interested reader to \cite{Boers2021CSDGreenlandIS,Bury2021CSDMachineLearning,Bury2023DiscreteBifML,Kuehn2013CSDVar,Dakos2012VarianceCanDecrease,Dakos2018MultDResiliencevsEigenvalues_EOF,morr2023internalnoiseinterferenceanticipation,donovan2022spatial,benmayi2024uncertainties, rietkerk2025ambiguity,Zhang2015CSDROC} and references therein for further discussion of EWS for tipping points.

There is some debate as to how much prediction CSD signals can show \cite{benmayi2024uncertainties, rietkerk2025ambiguity}. The danger is that by choosing observables, filtering schemes and/or thresholds it is possible to find accurate signals of historical tipping points but these may give false positives or false negatives when applied to predictions of future tipping points \cite{Ditlevsen2010SignalNoise, kefi2013early}. One could give up any claim of predictive ability of an EWS, but this somewhat defeats the utility of an ``early warning''. 

In this paper, we aim to bring this issue into focus by discussing factors needed for an EWS based on CSD to have good predictive skill of an approaching B-tipping. To make the problem tractable, we will specify a prediction time (time horizon) $\Delta>0$ and aim to make predictions forward over that time horizon, and for passage through fold bifurcations. For a prediction to be skilful, we claim that all of the following will be needed:
\begin{itemize}
    \item[(a)] Good separation of timescales between slow forcing and fast system timescales.
    \item[(b)] Good separation of nonlinear dynamics signal and low amplitude noise.
    \item[(c)] Reliable estimation of the stability of a system, and quantification of the uncertainty of that estimation.
    \item[(d)] Reliable estimation of an extrapolation timescale over which a significant trend in stability can be extrapolated into the future. 
\end{itemize}
Many of these points have been discussed in previous literature on the topic (notably (a-c) in \cite{Ditlevsen2023AMOCPrediction} and (d) in \cite{benmayi2024uncertainties}). In particular, (c) a very good fit to a trend, even over a long period, is no guarantee of (d) that the trend will extrapolate to a tipping point \cite{bastiaansen2023climate}. 

On a related topic, there is much interest in {\em cascades} of tipping elements \cite{dekker2018cascading,wunderling2024climate,ashwin2017fast,armstrong2022exceeding, wunderling2024climate}, where tipping of one system may influence tipping of another. In this context, we consider how a tipping event in an ``upstream'' system influences tipping of a ``downstream'' system within a cascade of tipping elements. Various studies have highlighted how upstream tipping can influence downstream tipping depending on the systems and their coupling, in particular \cite{Klose2021Cascade} and examples such as \cite{ritchie2025cascades,sinet2024amoc}.  In this paper, we focus on examples of what we call an {\em accelerating cascade of tipping elements}, where each system can potentially tip and the downstream system evolves on a faster timescale than the upstream system. Tipping of the upstream system can then be viewed as a slow change in the forcing of the downstream system. We find cases where early warning of tipping of the upstream system is possible, but early warning for the downstream system is not possible.

The paper is organised as follows. In Section~\ref{sec:setting}, we discuss an EWS for future tipping based on extrapolation of the trend of the return rate, as in \cite{Ditlevsen2023AMOCPrediction}, but limited to the finite time horizon so we can test the skill of the prediction. We discuss a method to estimate the skill of this test using the specified time horizon and a {\em receiver-operator characteristic/area under curve} (ROC/AUC) methodology \cite{fawcett2006introduction}. Section~\ref{sec:extrap} presents examples where this extrapolation is skilful for a range of time horizons, and examples where there are obstructions to extrapolation for various distinct reasons. In Section~\ref{sec:propagation} we consider an example of two coupled systems, each with a tipping point and with coupling between them. We state criteria such that an upstream system crossing a tipping point may (or may not) cause a downstream system to tip. We study a specific example of coupled one-dimensional systems to test how the coupling between systems affects tipping. Section~\ref{sec:EWS} examines a stochastic version of this coupled system and explores cases where skilful early warning for tipping of the downstream system is possible. We find that the downstream tipping may remain deeply unpredictable in cases of {\em downstream-within-upstream} tipping. Finally, in Section~\ref{sec:discuss} we discuss some implications and potential applications of this work.

\section{Timescales and tipping points}
\label{sec:setting}

In this section, we consider a nonlinear system whose state $x\in\R^n$ is governed by
\begin{equation}
\label{eq:f-original}
    \frac{d}{dt} x = f(x,\Lambda(rt))
\end{equation}
where $\Lambda(s)$ characterises the forcing input to the system, $r>0$ represents the rate of change of forcing and $f:\R^{n+p}\rightarrow \R^n$ is a smooth function. We assume that $\Lambda:\R\rightarrow \R^p$ is piecewise continuous.  In the limit where $r>0$ is small (and so $\Lambda$ is slowly varying) bifurcation-induced tipping points (B-tipping) of (\ref{eq:f-original}) can be understood using methods discussed in \cite{Ashwin2012tipTypes,ashwin2017parameter,wieczorek2023rate}; i.e. tipping in (\ref{eq:f-original}) can be understood in terms of  bifurcations of attractors of the associated {\em frozen system}
\begin{equation}
\label{eq:f-frozen}
    \frac{d}{dt} x = f(x,\lambda)
\end{equation}
on varying $\lambda\in\R^p$ and for convenience we suppose that 
$$
\Lambda(s)\rightarrow\lambda_{\pm}~\mbox{ as }s\rightarrow \pm\infty
$$
so there are limiting autonomous past and future systems.
If $\Lambda(s)$ remains in a region of parameter space where there is a branch of stable equilibria for (\ref{eq:f-frozen}) then, for small enough $r>0$, solutions will {\em track} the branch of equilibria. On the other hand, if $\Lambda$ crosses a fold (saddle-node) bifurcation where a stable equilibrium ceases to exist then this implies solutions will undergo B-tipping. 

We assume for small enough $r>0$ there is a (local) pullback attractor $x^{[r]}(t)$ of (\ref{eq:f-original}) that limits to $X_-$ in the past and $X_+$ in the future  \cite{ashwin2017parameter}, for some attractors $X_{\pm}$ of (\ref{eq:f-frozen}) with $\lambda=\lambda_{\pm}$. We say there is {\em end-point tracking} of an attractor from $X_-$ to $X_+$ if there is a branch of attractors $X(\Lambda(s))$ of the frozen system such that 
$$
X_-=X(\lambda_-),~~X_+=X(\lambda_+).
$$
{\em Bifurcation induced tipping} (B-tipping) occurs if for all small enough $r>0$ there is no such branch that connects $X_-$ and $X_+$, and (in the limit of small $r$) occurs when $\Lambda(rt)$ crosses some bifurcation point $\lambda_f$.

Noise added to a system such as (\ref{eq:f-original}) can give rise to {\em early warning signals} (EWS) or {\em precursors} of B-tipping because the response to noise gives information about the stability of equilibria of the noise-free system. Consider a stochastic version of (\ref{eq:f-original}): 
\begin{equation}
\begin{aligned}
dx &= f(x,\Lambda(rt))dt + \sigma dW_t
\end{aligned}
\label{eq:f-nonaut-sde}
\end{equation} 
where $\sigma$ is small, $W_t$ is an $\R^n$-valued standard Wiener process with covariance 
$$
\bE((W_t-W_0)^T(W_t-W_0))=t\Sigma
$$
for some positive definite $\Sigma$ and suppose that (\ref{eq:f-frozen}) has a branch of stable equilibria.

\subsection{Early warning prediction}\label{sec:EWSskill}

The goal of EWS is to predict future tipping from data up to time $t = t_p$. To make this more tangible, we focus here on predicting whether tipping is going to happen within some specific time horizon i.e. between $t = t_p$ and $t = t_p + \Delta$ for some $\Delta>0$.

To quantify prediction skill, we need a clear notion of whether a system tips. We say a \emph{tipping event is underway} at time $t$ if some chosen (derived) scalar property or observable $\Psi$ of the system passes a chosen threshold $\theta$.\footnote{$\Psi(x(t))$ is typically the rate of change of some observation of the system at time $t$, and $\theta$ a threshold above which we consider the system to be in the process of tipping.} Using such a definition, we compute the minimum value $\delta$ such that $\Psi(x(t_p+\delta)) \geq \theta$. If this value is below $\Delta$, that means tipping is happening in the time horizon of interest. This can be expressed as an indicator function $I_{\tip,\Delta}(t_p)$ that returns $1$ if a tipping happens in this time horizon and $0$ otherwise, i.e.
\begin{equation}
I_{\tip,\Delta}(t_p) := \bI_{\min\{s\geq 0~:~\Psi(x(t_p+s))\geq \theta\} \in [0,\Delta]}
\end{equation}

An {\em early warning predictor} is a function $I_{\EWS,\Delta}(t_p)\in\{0,1\}$ derived from the timeseries up to prediction time $t_p$ that is a predictor for $I_{\tip,\Delta}(t_p)$. We derive such a predictor using a {\em scoring classifier} \cite{fawcett2006introduction};\footnote{Note we take the convention that a low value of the predictor suggests a prediction of tipping.} this is a real-valued function $C_{\Delta}(t_p)$. Given threshold $\kappa$ we consider
$$
I_{\EWS,\Delta,\kappa}(t_p) :=\bI_{C_{\Delta}(t_p)\leq \kappa}
$$
as an estimator for $I_{\tip,\Delta}(t_p)$. The threshold $\kappa$ can be chosen to optimise the skill of the predictor by balancing the ability to correctly detect negatives or positives. Note that for any $\kappa<\kappa'$ we have
$$
I_{\EWS,\Delta,\kappa}(t_p)\leq I_{\EWS,\Delta,\kappa'}(t_p).
$$
Moreover, note that for all $\Delta>0$ and $t_p$, and almost all realizations we will have
\begin{equation}
    \lim_{\kappa\rightarrow -\infty} I_{\EWS,\Delta,\kappa}(t_p)=0,~~
   \lim_{\kappa\rightarrow +\infty} I_{\EWS,\Delta,\kappa}(t_p)=1
    \label{eq:predictlimits}
\end{equation}
i.e. there are limiting thresholds where $I_{\EWS,\Delta,\kappa}$ systematically predicts no tipping or inevitable tipping. 

The prediction will be maximally skilful if there is a choice of threshold that minimises false positives and false negatives. For some test ensemble we can quantify the proportion of false positives $FP$ and true positives $TP$ for that ensemble as a function of the threshold $\kappa$ using an ROC/AUC analysis \cite{fawcett2006introduction}. This approach has been used to understand skill of EWS for example in:  \cite{Proverbio2022COVIDEWS,Brett2020ROCEWS,donovan2022spatial,Zhang2015CSDROC}; see \cite{Lehnertz2024timeseries} for a comprehensive review with particular emphasis on surrogate methods.

Note that the ROC goes from $(FP,TP)=(0,0)$ to $(1,1)$ on varying $\kappa$; for too low $\kappa$ the predictor always predicts no tipping, whereas for too high $\kappa$ the predictor always predicts tipping. The {\em optimal choice for the threshold} is the value $\kappa=\kappa^{Opt}(t_p,\Delta)$ that corresponds to the closest point to $(0,1)$, i.e. that minimises both false negatives and false positives. In general this will depend on $t_p$, $\Delta$ and the ensemble considered. How optimal this choice can be made is expressed by the {\em area under the curve} (AUC) for this ROC \cite{fawcett2006introduction,Lehnertz2024timeseries}. The closer that AUC is to one, the more skilful a test can be made by judicious choice of $\kappa$. This will clearly depend on the ensemble, the quality of the estimator and the quality of the fit to a trend, as well as prediction time $t_p$ and time horizon $\Delta$.

\subsection{Estimating stability from critical slowing down}
\label{sec:estimation}

In this section, we consider an explicit early warning predictor based on return rate extrapolation. For given $\lambda$, suppose there is a stable equilibrium of (\ref{eq:f-frozen}) at $x^*(\lambda)$. For $x$ close to $x^*$ we can approximate the stochastic parameter frozen version of (\ref{eq:f-nonaut-sde}), namely
\begin{equation}
dx=f(x,\lambda)dt+\sigma dW_t
\end{equation}
as a linear SDE whose solutions are (multi-dimensional) Ornstein-Uhlenbeck processes
\begin{equation}
 dx = Df(x^*,\lambda) (x-x^*) dt+\sigma dW_t.	
\end{equation}
If we assume there is a unique least stable (or {\em leading}) eigenvalue of $Df(x^*,\lambda)$, namely $-\alpha$ with $\alpha>0$, then we say $\alpha$ is the {\em return rate}. When this approaches zero, that is a sign of loss of stability \cite{wissel1984universal}. There is locally a projection of $x$ onto an SDE for $z\in\R$ that, if linearised, has the form
\begin{equation}
dz = -\alpha (z-\mu) dt+\sigma dV_t	
\label{eq:OU-1d}
\end{equation}
for some Wiener process $V_t$. A more general observable may include noise from other components \cite{morr2023internalnoiseinterferenceanticipation}, but we shall ignore this issue here.

We can estimate $\alpha$ from timeseries $z(t)$ at some prediction time $t_p$ by computing the AR1 coefficient $\rho$ of $z_i=z(t_i)$ in an observation window at times $\{t_i\}_{i=0}^{n_w}$ with $t_{i}=t_p+(i-n_w)\delta$. Namely we fit to
\begin{equation}
z_{i+1}=\rho (z_i-\mu) +\sigma e_i
\end{equation}	
for $i=0,\ldots,n_w$
where $e_i$ is normally distributed and $0<\rho<1$ corresponds to $\rho=e^{-\alpha\delta}$. For $\rho$ close to 1, maximum likelihood estimators (MLE) for $\mu$ and $\rho$  \cite{Ditlevsen2023AMOCPrediction} are:
\begin{equation}
\hat{\mu} = \frac{1}{n_w+1}\sum_{i=0}^{n_w} z_i,
~~~~\hat{\rho} = \frac{\sum_{i=1}^{n_w} (z_{i-1}-\hat{\mu})(z_i-\hat{\mu})}{\sum_{i=0}^{n_w} (z_i-\hat{\mu})^2}.
\end{equation}
An estimator for the return rate $\alpha>0$ is therefore
\begin{equation}
\hat{\alpha}=-\ln(\hat{\rho})/\delta.
\end{equation}
In the case $0<\alpha\delta\ll 1$ the standard deviation of the estimator $\hat\rho$ can be approximated as $\hat{\sigma}= \sqrt{2\hat{\rho}\delta/n_w}$, as in \cite{Ditlevsen2023AMOCPrediction}. Hence an estimator of two standard deviations of $\alpha$ is given by $\hat{\alpha}_{\pm}=-\ln(\hat{\rho}\mp 2\hat{\sigma})/\delta$. 

\subsection{Early warning prediction from extrapolation}
\label{sec:class}

In the previous subsection, we have obtained an estimator $\hat{\alpha}$ based on an observation window. If the system is non-autonomous but slowly varying (i.e. the rate of change is slower than that observation window length) then we obtain a time-dependent estimator $\hat{\alpha}(t)$. In this subsection we show how to do this by extrapolating $\hat{\alpha}(t)$ to obtain the family of early warning predictors $I_{\EWS,\Delta,\kappa}(t)$.

We assume linear drift of a parameter through a fold (saddle-node) bifurcation\footnote{See Section~\ref{sec:discuss} for discussion of other bifurcation types.}, as discussed in \cite{Ditlevsen2023AMOCPrediction}.
For transverse passage through a fold bifurcation, (\ref{eq:f-nonaut-sde}) is locally topologically equivalent to a normal form
\begin{equation}\label{eq:normal_form}
f(x,\lambda)=a_0 x^2+ a_1 \lambda
\end{equation}
where $a_0\neq 0$ and $a_1\neq 0$. In this case there is a near identity transformation that removes higher order terms in $x$ and $\lambda$ \cite{Kuznetsov2004AppliedBifurcationTheory}. If the parameter passes through the fold bifurcation with non-zero speed at $t=t_0$ then we have (\ref{eq:f-nonaut-sde}) with
$$
f(x,\Lambda(rt))=a_0 x^2+ \tilde{a}_1 (t-t_0)
$$
for some $\tilde{a}_1\neq 0$. We assume $a_0\tilde{a}_1<0$ so that, in the absence of noise, the quasistatic system has an equilibria at $x_{\pm}= \pm\sqrt{(t-t_0)\tilde{a}_1/a_0}$ with eigenvalue $\alpha(t)=Df(x_{\pm})= \pm 2 a_0 \sqrt{(t-t_0)\tilde{a}_1/a_0}$ for $t<t_0$.
This suggests (see eg \cite{Ditlevsen2023AMOCPrediction}) that near the fold bifucation we can expect a linear relationship between time and the square of the eigenvalue $\alpha^2$
$$
\alpha^2(t)=4 |a_0\tilde{a}_1| (t_0-t).
$$
Hence \cite{Ditlevsen2023AMOCPrediction} propose a linear fit of $\alpha^2$ over the period $t\in[t_p-n_2\delta,t_p]$ to obtain
\begin{equation}
\hat{\alpha}^2_{t_p}(t)\approx \beta_0(t_p)+\beta_1(t_p)(t-t_p).
\label{eq:alphasqfit}
\end{equation}
This gives values for $\beta_i(t_p)$ where $\beta_0(t_p)\geq 0$ and $\beta_1(t_p)>0$ (resp. $\beta_1(t_p)<0$) corresponds to a trend\footnote{Another popular method to detect a trend for EWS is the Kendall $\tau$ statistic, though this does not easily lead to a predictor in our sense.} with increasing (resp. decreasing) stability.

One can extrapolate (\ref{eq:alphasqfit}) to predict a tipping point such that $\hat{\alpha}^2_{t_p}(t_{\tip})=0$ which will be at time 
\begin{equation}
t_{\tip}=t_p-\beta_0(t_p)/\beta_1(t_p)
\end{equation}
This is in the future if and only if $\beta_1(t_p)<0$ i.e. if there is a decreasing trend. This can be used to infer a distribution of future tipping times $t_{\tip}$ as in \cite{Ditlevsen2023AMOCPrediction}. However, it is hard to judge the skill of this in cases $t_{\tip}>t_p+\Delta$ where the errors may be dominated by extrapolation.

\begin{figure}
    \centering
    \includegraphics[width=0.7\linewidth]{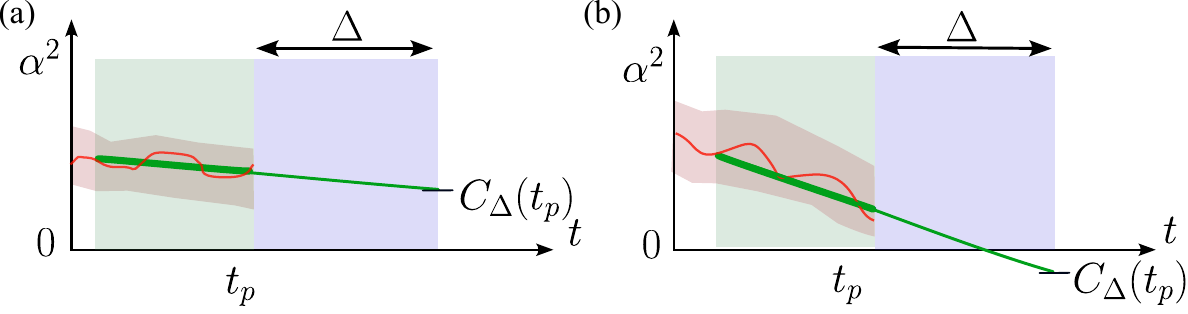}
    \caption{Construction of the scoring classifier $C_{\Delta}(t_p)$ to predict a tipping point in time horizon $\Delta$ from a prediction time $t_p$, assuming proximity to a fold bifurcation. The red line is the estimator of $\alpha^2(t)$ with confidence interval (light red) up to prediction time $t_p$. Extrapolating from the green region to $t_p+\Delta$, gives a classifier that predicts tipping occurs if it is below some threshold.  For threshold $\kappa$ we predict tipping in the blue region $[t_p,t_p+\Delta]$ if $C_{\Delta}(t_p)<\kappa$. In case (a) the extrapolation is above zero meaning $\kappa$ needs to be sufficiently positive to predict tipping, while in case (b) tipping will be predicted for  all $\kappa$ larger than some negative quantity.}
    \label{fig:ews-extrapolation}
\end{figure}

This suggests the scoring classifier $C_\Delta(t_p)$ (see Section~\ref{sec:EWSskill})
\begin{equation}
C_{\Delta}(t_p):=\beta_0(t_p)+\beta_1(t_p)\Delta.
\label{eq:CDelta}
\end{equation}
The construction of this classifier is shown in Figure~\ref{fig:ews-extrapolation}. We use $C_{\Delta}(t_p)$ to define a family of early warning predictors:
\begin{equation}
    I_{\EWS,\Delta,\kappa}(t_p) = \bI_{C_{\Delta}(t_p)\leq \kappa}.
    \label{eq:predict}
\end{equation}
Note that the predictor (\ref{eq:predict}) is monotonic in $\kappa$ and has limits as in (\ref{eq:predictlimits}). Hence we can use the ROC/AUC methodology outlined at the end of Section~\ref{sec:EWSskill} to quantify the skill of the predictor.

\section{Predictive skill and extrapolation problems}
\label{sec:extrap}

In Section~\ref{sec:exampleROCAUC} we first verify, using the classifier from Section~\ref{sec:class} and an ROC/AUC methodology, that an early warning predictor based on this classifier can have good skill in case of a monotonic passage through a fold bifurcation. However, the remainder of the section highlights three independent reasons for the breakdown of extrapolation -- and thus the loss of skill -- via examples. Section~\ref{sec:examplenonmon} highlights a problem due to non-monotonic forcing in time, Section~\ref{sec:examplevalidity} highlights a problem due to predicting outside a region of validity of a normal form; Section~\ref{sec:exampleeigenvalue} highlights a problem due to existence of other stable eigenvalues. 

\subsection{Example of quantification of skill using ROC/AUC}
\label{sec:exampleROCAUC}

\begin{figure}
\centering
	\includegraphics[width=9cm]{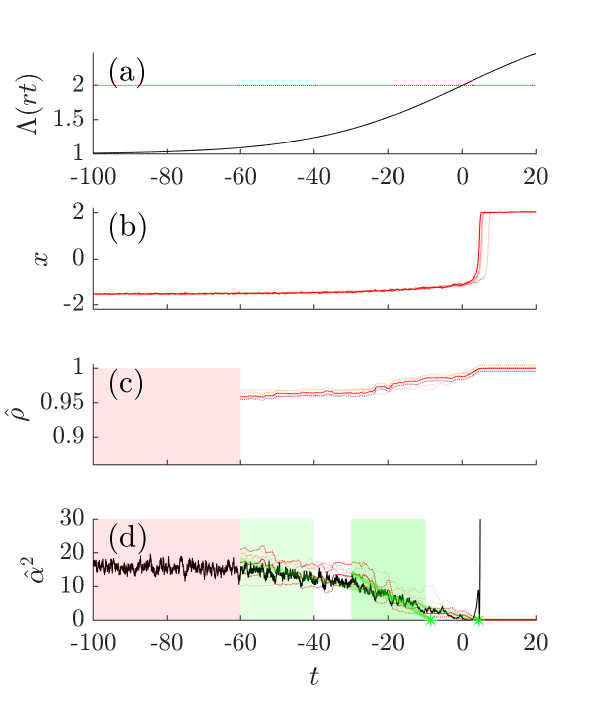}
\caption{Slow monotonic passage through a tipping point with a parameter shift (\ref{eq:tanhbump}) from $\lambda_-=1$ to $\lambda_+=3$ for system (\ref{eq:examplenonmon}). (a) The forcing $\Lambda(rt)$ crosses a fold bifurcation at $\lambda=2$ at $t = 0$. (b) Example trajectory $x(t)$ (bold) from an ensemble of noise realizations (faint) sampled with timestep $\delta$. (c) AR1 estimator $\hat{\rho}$; the light red region represents the length of window used to detrend and compute the estimators. (d) Red: estimator of return rate $\hat{\alpha}=-\ln (\hat{\rho})/\delta$, with dashed lines showing two standard deviations of the estimator (dashed lines) and the remaining realizations (faint). The black line shows $\alpha^2$ where $\alpha=-Df(x(t),\Lambda(t))$ is the (actual) return rate. The estimator is fitted to $\hat{\alpha}^2_{t_p}(t)\approx \beta_0+\beta_1(t-t_p)$ in the light green regions up to times $t_p=-40$ and $t_p=-10$ and extrapolated forwards; note the relatively accurate extrapolation (but significant variability) of the fits in this case.
}
\label{fig:ews-numerics-cases-shift}
\end{figure}

Consider a system with state $x\in\R$ governed by
\begin{equation}
    \begin{aligned}
        dx =& f(x,\Lambda(rt))dt+\eta dW_t
    \end{aligned}
    \label{eq:examplenonmon}
\end{equation}
where $f(x,\lambda) : = 3x-x^3+\lambda$. This has a region of bistability for $\lambda\in [-2,2]$ with fold bifurcations at $\lambda_u=2$ and $\lambda_l=-2$. We consider a family of monotonic parameter shifts of the form
\begin{equation}
\begin{aligned}
\Lambda(s) &: = \lambda_-+\frac{1}{2}(\lambda_+-\lambda_-)\left(\tanh(s)+1\right).
\end{aligned}
\label{eq:monotonic}
\end{equation}
We fix parameters
\begin{equation}
\lambda_-=1,~\eta=0.05,~r=0.025.
\label{eq:params-default}
\end{equation}
The other parameter ($\lambda_+$) will be varied in the ensemble considered for ROC/AUC analysis. Note that if $\lambda_+ > 2$ then (in the absence of noise) B-tipping occurs, whereas there is end-point tracking for $\lambda_+ < 2$. In Figure~\ref{fig:ews-numerics-cases-shift}, an example trajectory, including stability estimation and early warning predictor, is given in case of B-tipping occurring, i.e. with $\lambda_+ = 3$.

Figure~\ref{fig:ews-numerics-cases-shift} shows: (a) the time variation of $\Lambda(rt)$; (b) a number of realizations starting at time $t=-100$ close to the lower branch; (c) an estimator $\hat{\rho}$ for the AR1 coefficient up to prediction time $t_p=-40$ or $t_p = -10$; (d) the estimator $\hat{\alpha}^2$ for the return rate (red), and the actual value of the Jacobian at the current trajectory position (black). In (c-d) the pink window shows the length of the moving window over which detrending and estimation take place. The light green window is used to fit a linear passage through a fold bifurcation. The green line shows the best fit in this window, extrapolated into the future. The value of this extrapolation at time $t_p+\Delta$ is the scoring classifier as shown in Figure~\ref{fig:ews-extrapolation} and discussed in Section~\ref{sec:class}.

\begin{figure}
    \centering
    \includegraphics[width=\textwidth]{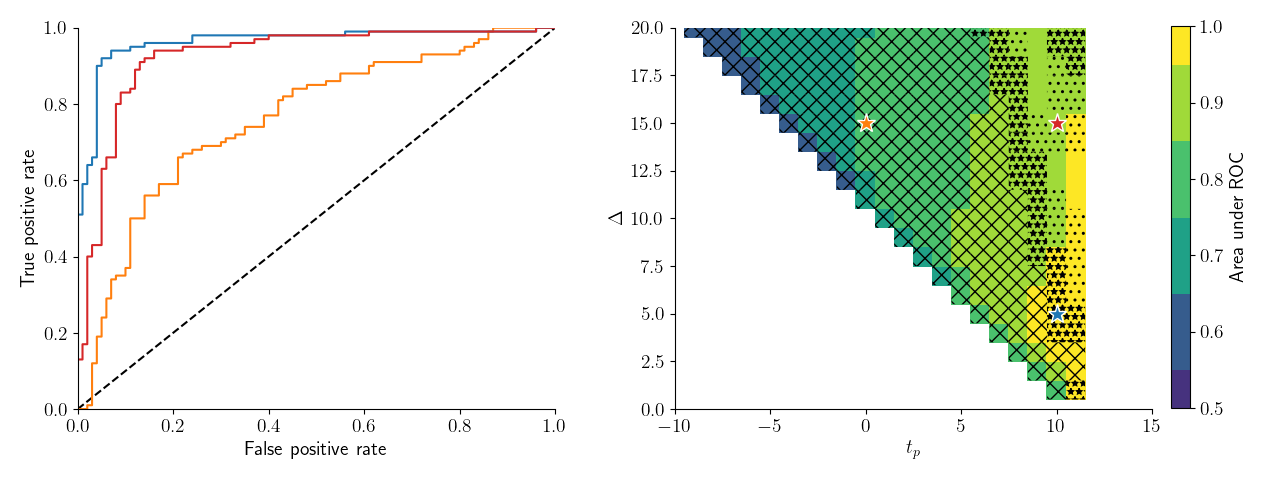}
    
    \caption{Quantifying skilful prediction at time $t_p$ for system \eqref{eq:examplenonmon} forced by the monotonic parameter shift (\ref{eq:monotonic}) tipping within the window $[t_p,t_p+\Delta]$. This corresponds to the example in Section~\ref{sec:exampleROCAUC} where $\lambda_+$ is chosen randomly with equal probability from $\{1,3\}$.
    Left: ROC curves for prediction times $t_p$, and time horizons $\Delta$ indicated by the coloured stars in the right panel. Right: Colour plot of area under ROC curve (AUC) for different prediction times, $t_p$ and time horizons $\Delta$, using a fitting window of length 10. To construct the curves, 1,000 simulations are performed (with
    a forcing parameter randomly sampled).
    Then, for each prediction time $t_p$ and horizon $\Delta$, out of those 1,000 simulations, an ensemble was created by picking the first 100 that show tipping and the first 100 that do not show tipping within the time window $[t_p, t_p+\Delta]$. If such an ensemble cannot be created, no ROC/AUC analysis was performed (white areas in right panels). The ROC is calculated using these simulations and systematically considering a sufficiently wide range of thresholds, $\kappa$, such that for the maximum $\kappa$ all trajectories are predicted to tip and the minimum $\kappa$ all trajectories are predicted not to tip. 
    The hatching denotes the optimal threshold, $\kappa^{Opt}(t_p,\Delta)$: Cross hatching,  $|\kappa^{Opt}(t_p,\Delta)|>10$; star hatching, $5\leq |\kappa^{Opt}(t_p,\Delta)|<10$; dot hatching, $2\leq |\kappa^{Opt}(t_p,\Delta)|<5$; no hatching, $|\kappa^{Opt}(t_p,\Delta)|<2$; see text for more details. Note there is good skill close enough to the tipping but further in the past the predictive power drops off quite rapidly.
    }
    \label{fig:roc_auc_upstream}
\end{figure}

The upper panels of Figure~\ref{fig:roc_auc_upstream}
correspond to a ROC/AUC analysis of an ensemble of such monotonic forcing scenarios. The ROC/AUC are computed for this example using an ensemble of realizations where $\lambda_+$ is chosen randomly and i.i.d. from $\{1,3\}$, $\lambda_-=0$, $r=0.05$ and other parameters as in (\ref{eq:params-default}). To construct the curves, 1,000 simulations were performed (with $\lambda_+$ randomly sampled). Then, for each prediction time $t_p$ and horizon $\Delta$, out of those 1,000 simulations an ensemble was created by picking the first 100 that show tipping and the first 100 that do not show tipping within the time window $[t_p, t_p+\Delta]$. If such ensemble cannot be created, no ROC/AUC analysis was performed. The left column shows examples of ROC curves computed for three cases of prediction time $t_p$ and horizon $\Delta$ corresponding to the starred locations in the right panel. The plot on the right shows AUC computed in the coloured region where there are a sufficient number of tipping events in $[t_p,t_p+\Delta]$ (in order to be able to compute statistics for false positives/negatives properly). Observe that $t_p$ needs to be close to the threshold crossing if there is one, for the AUC to be computable -- but good skill can definitely be achieved by choosing the threshold $\kappa$ careful when $\Delta$ is small and $t_p$ close to the threshold crossing. However, the skill quickly drops off as $\Delta$ becomes larger or $t_p$ is further away from the threshold crossing, even in this idealised case.

\subsection{Example: lack of extrapolation due to non-monotonic forcing}
\label{sec:examplenonmon}

\begin{figure}
\centering
    \includegraphics[width=9cm]{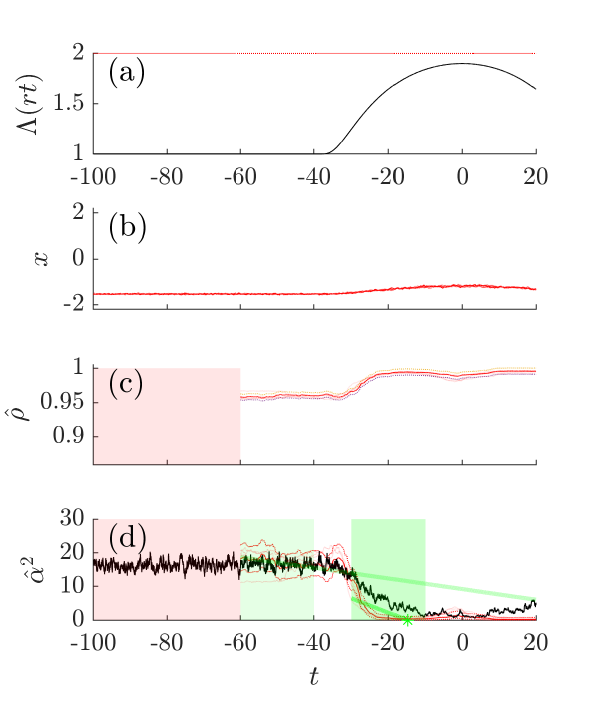}
\caption{As in Figure~\ref{fig:ews-numerics-cases-shift} except $\lambda_+=1$ and $\tilde{\lambda}=1.9$, giving non-monotonic forcing, showing extrapolations from $t_p=-40$ and $t_p=-10$. In this case, there is no tipping - but extrapolation from $t_p=-10$ predicts tipping even though the non-monotonic forcing does not cross the threshold $\lambda=2$.
}
\label{fig:ews-numerics-cases-bump}
\end{figure}

Next, we turn to cases in which the early warning prediction might break down. For this, we first consider the same system \eqref{eq:exampleleading}, but with non-monotonic parameter shifts of the form:
\begin{equation}
\begin{aligned}
\Lambda(s) &: = \lambda_-+\frac{1}{2}(\lambda_+-\lambda_-)\left(\tanh(s)+1\right)+\tilde{\lambda}\bump(as+b).
\end{aligned}
\label{eq:tanhbump}
\end{equation}
where we define
\begin{equation*}
    \bump(s) = \left\{
    \begin{array}{cl}
    \exp \left[\frac{x^2}{x^2-1}\right] & \mbox{ if } |x|<1\\
    0 & \mbox{ otherwise}.
    \end{array} \right. \label{eq:bump}
\end{equation*}
By choosing the coefficients in (\ref{eq:monotonic}), we can explore paths with limits $\lambda_{\pm}$ as $t\rightarrow \pm\infty$ that may be non-monotonic if $\tilde{\lambda}\neq 0$.  Figure \ref{fig:ews-numerics-cases-shift} shows a monotonic parameter shift with $\lambda_+ = 3$ and $\tilde{\lambda} = 0$ that lead to B-tipping. By contrast, Figure \ref{fig:ews-numerics-cases-bump} shows examples with this non-monotonic forcing with $\lambda_+ = 1$, $\tilde{\lambda} = 1.9$, $a = 1$ and $b = 0$ (other parameters as in \eqref{eq:params-default}). This gives a parameter shift that comes close to, but avoids, crossing $\lambda_u = 2$ -- and hence (in the absence of noise) no B-tipping occurs.
For Figure~\ref{fig:ews-numerics-cases-shift} the decreasing trend in $\alpha$ is identified, predicting a tipping point close to the actual tipping event. By contrast, Figure~\ref{fig:ews-numerics-cases-bump} shows that a decreasing trend in $\alpha$ is correctly identified, but extrapolation close to the minimum yields a prediction of tipping although no tipping occurs in this case.

\begin{figure}
    \centering
    \includegraphics[width=\textwidth]{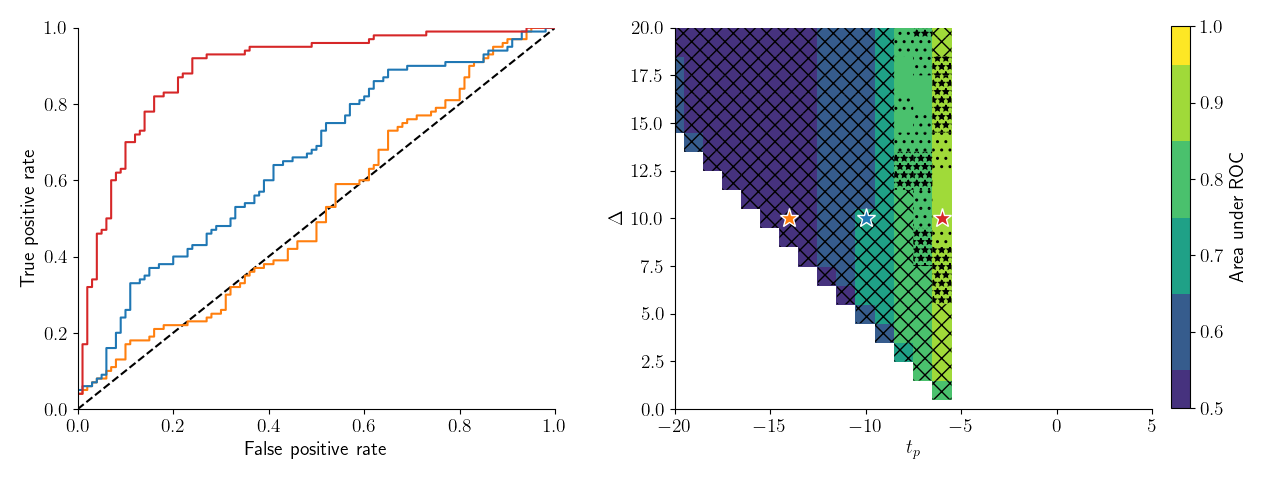}    
    \caption{Quantifying skilful prediction at time $t_p$ for system \eqref{eq:examplenonmon} forced by the nonmonotonic parameter dependence (\ref{eq:tanhbump}) tipping within the window $[t_p,t_p+\Delta]$, as in Figure~\ref{fig:roc_auc_upstream}. This corresponds to the example in Section~\ref{sec:examplenonmon} where $\tilde\lambda$ is chosen randomly with equal probability from $\{1.5,2.5\}$ and $\lambda_-=\lambda_+=0$, $a=0.05$, $b=0$ and using a fitting window of length 10.
    Left: ROC curves for prediction times $t_p$, and time horizons $\Delta$ indicated by the coloured stars in the right panel. Right: Colour plot of area under ROC curve (AUC) for different prediction times, $t_p$ and time horizons $\Delta$. Observe in this case the skill is substantially lower than the case in Figure~\ref{fig:roc_auc_upstream} even for the most optimal choices of $t_p$ and $\Delta$, due to failure of validity of the extrapolation.
    }
    \label{fig:roc_auc_nonmon}
\end{figure}

Figure~\ref{fig:roc_auc_nonmon} shows a ROC/AUC analysis of an ensemble of
such non-monotonic forcing scenarios. To construct these ROC curves, we take $\lambda_- = \lambda_+ = 0 < \lambda_u = 2$, $a = 0.05$ and $b = 0$, and chosen the bump strength $\tilde{\lambda}$ uniformly from $\{1.5,2.5\}$. For some of the ensemble $\lambda_u = 2$ is crossed and B-tipping does happen. The left column show examples of ROC curves computed for three different prediction times $t_p$ and $\Delta$ corresponding to the starred locations in the right panel.  It can be seen that the predictive skill for the early warning predictor is less for the non-monotonic forcing ensemble even for optimal choice of threshold. In particular, it shows that all predictive skill can be lost when $t_p$ is too far before a potential tipping and/or $\Delta$ is too large. This indicates that in such non-monotonic forcing scenarios, an early warning predictor may need to be hyper-tuned to have good prediction skill -- and even then prediction can only be expected close to a potential tipping.

\subsection{Example: lack of extrapolation due to being outside region of validity of normal form}
\label{sec:examplevalidity}

\begin{figure}
    \centering
    \includegraphics[width=0.5\linewidth]{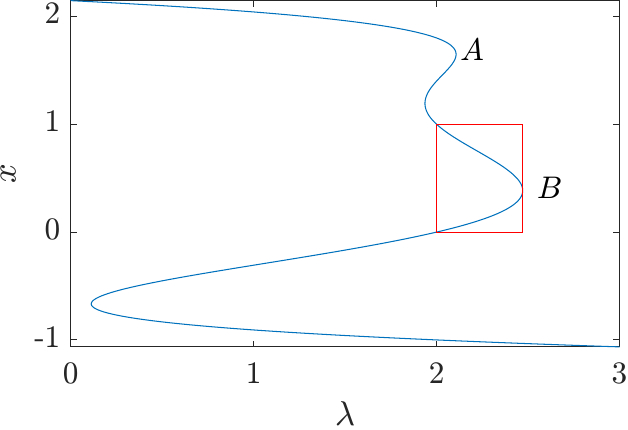}
    \caption{Equilibria for the system (\ref{eq:Feq},\ref{eq:F}) with $\eta=0$. On increasing $\lambda$ in this region and starting on the upper branches we must pass fold bifurcations $A$ and $B$ causing B-tipping on slowly increasing $\lambda$. The red rectangle outlines a region where the number of branches are topologically equivalent to the normal form of the fold bifurcation $B$. Note that no extrapolation from the upper branch, i.e. up to fold $A$, can be expected to be an accurate predictor for fold $B$.}
    \label{fig:fifthorder}
\end{figure}

The extrapolation step might also break down when the assumption of the normal form \eqref{eq:normal_form} does not hold. To illustrate this, consider the system
\begin{equation}
    \begin{aligned}
        dx =& (F(x)-\Lambda(rt))dt+\eta dW_t
    \end{aligned}
    \label{eq:Feq}
\end{equation}
where $F(x)$ is a smooth function and $\Lambda$ monotonically increases from $\lambda_-$ to $\lambda_+$. The turning points of $F$ (i.e where $DF(x_0)=0$) correspond to bifurcations of the  noise-free ($\eta=0$) system when $\Lambda=-F(x_0)$. For example, Figure~\ref{fig:fifthorder} shows a bifurcation diagram for 
\begin{equation}
F(x)=-x(x^2-1)(x-1.4)(x-1.8)+2
\label{eq:F}
\end{equation}
which has four fold bifurcations in the region illustrated, two of which will be responsible for B-tipping as $\lambda$ increases. Indeed, for monotonic increase of $\lambda$, a trajectory starting on the upper branch will pass through two fold bifurcations $A$ and then $B$ in sequence. The red rectangle approximates a region of validity of the normal form for the fold $B$. Note that trajectories can only enter the red region after passing fold $A$ and so trends from the upper branch cannot be extrapolated to reliably give a prediction of passing the fold $B$.

\subsection{Example: lack of extrapolation due to change of leading eigenvalue}
\label{sec:exampleeigenvalue}

\begin{figure}
\centering
    \includegraphics[width=9cm]{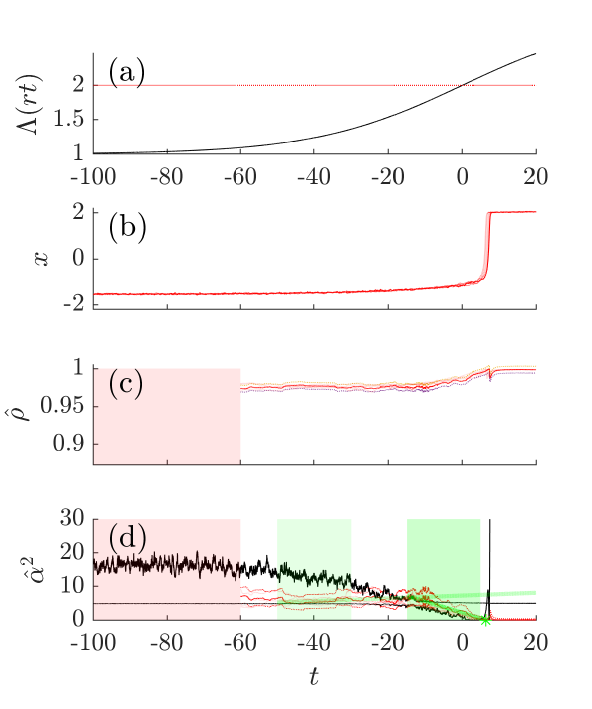}
\caption{As in Figure~\ref{fig:ews-numerics-cases-shift} for the fold normal form with an additional stable direction (\ref{eq:exampleleading}), extrapolating from two intervals up to prediction times $t_p=-40$ and $t_p=5$. We plot both eigenvalues of $Df(x(t))$ in black (note both are real for this example). We estimate $\hat{\alpha}^2$ (red) from projecting the timeseries onto the leading eigendirection at prediction time. Up to time $t\approx -30$ the leading eigenvalue is $\alpha^2 \approx 8$ with little trend. After this there is an avoided crossing of eigenvalues near $t=-30$ and the trend to tipping is more accurately captured.  
}
\label{fig:ews-leadingeig}
\end{figure}

Consider an example where $(x,y)\in\R^2$ is governed by
\begin{equation}
    \begin{aligned}
        dx =& (f(x,\Lambda(rt))+\xi_{12}y)dt+\eta_1 dV_t\\
        dy =& (\xi_{21}x+\xi_{22}y)dt+\eta_2 dW_t\\
    \end{aligned}
    \label{eq:exampleleading}
\end{equation}
where $f(x,\lambda) : = 3x-x^3+\lambda$, $V_t$ and $W_t$ are independent Brownian processes, parameters otherwise as in \eqref{eq:monotonic}-\eqref{eq:params-default} and additionally we set
$$
\xi_{12}=\xi_{21}=0.2,~\xi_{22}=-\sqrt{8},~\eta_1=\eta_2=0.05.
$$
A steady state of this system has two eigenvalues. One of them is close to $-8$ for all $\lambda$ values, whereas the second one organises the fold bifurcation; this second one is more negative than $-8$ unless $\lambda$ is close to the threshold crossing. Hence, if we project $(x,y)$ onto the leading eigendirection at the end of each window, and find the return rate from the AR1 coefficient, we typically find the eigenvalue that does not organise the fold bifurcation. Specifically, as shown in Figure~\ref{fig:ews-leadingeig} the presence of a second eigenvalue $\alpha^2\approx 8$ means that the least stable eigenvalue has no significant trend until shortly before the tipping point, suggesting extrapolation is only skilful close to tipping.

\section{Extrapolation and tipping cascades}
\label{sec:propagation}

\begin{figure}
    \centering
    \includegraphics[width=0.7\linewidth]{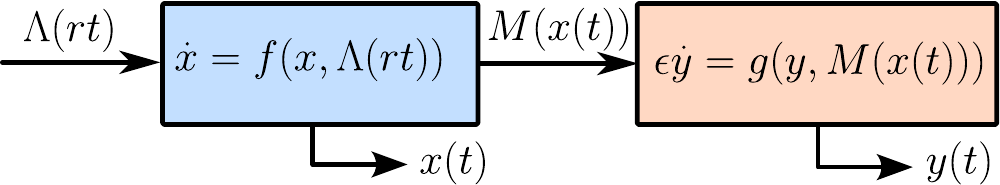}
    \caption{Block diagram showing a slowly varying input $\Lambda(rt)$ forcing an upstream nonlinear system with state $x(t)$  coupled to a  downstream system with state $y(t)$. We consider the case where $r>0$ and $\epsilon>0$ are both small meaning we have very slow forcing of a slow system that then forces a faster system. We call this an accelerating cascade of tipping elements.}
    \label{fig:blockdiag}
\end{figure}

In this section, we turn our attention to tipping cascades and their predictability. We focus on a simple case of an upstream system coupled to a downstream system, with external forcing and timescale separation. We suppose that, as summarised in a block diagram in Figure~\ref{fig:blockdiag}, there is a slowly forced {\em upstream (sub)system} for $x(t)$
\begin{equation}
\frac{d}{dt} x = f(x,\Lambda(rt))
\label{eq:f-nonaut}
\end{equation}
and this forces a {\em downstream (sub)system} with state $y\in\R^m$ governed by
\begin{equation}
\epsilon\frac{d}{dt} y = g(y, M(x(t)) ).
\label{eq:g-nonaut}
\end{equation}
We consider first the behaviour of the deterministic dynamics before introducing stochastic forcing in Section~\ref{sec:EWS}.
Note that $r$ is the rate of forcing, $\epsilon$ is the ratio of timescales of $x$ and $y$, $M:\R^n\rightarrow \R^q$ and $g:\R^{m+q}\rightarrow \R^m$ is smooth. Note also that (\ref{eq:g-nonaut}) is forced by a function of the output from the upstream system.  We assume $r>0$ is small (the system is slowly forced) and $\epsilon$ is small (the downstream dynamics evolves on a faster timescale than the upstream) and say the cascade of tipping elements {\em accelerates}.\footnote{For the case of a {\em decelerating} cascade ($\epsilon\gg 1$), it is a moot point whether the downstream system can be considered to B-tip. In this case, the changes to forcing $M(x(t))$ of the downstream timescale may not be small or slow.} See \cite{ritchie2025cascades} for discussion of the deterministic dynamics in more detail.

For an accelerating cascade, tipping of the upstream system $x$ may or may not cause a tipping of the fast downstream system $y$, depending on properties of $M$ and the tipping trajectory of $x$.
Writing the downstream system in the downstream timescale $\tau=t/\epsilon$ we see that $y(\tau)$ satisfies:
\begin{equation}
\frac{d}{d\tau} y = g(y,M(x(\epsilon \tau)) ).
\label{eq:g-nonaut-tau}
\end{equation}
One can understand the behaviour of (\ref{eq:g-nonaut-tau}) in relation to the upstream forcing $M(x(\epsilon \tau))$ and the {\em frozen downstream system}
\begin{equation}
\frac{d}{d\tau}y = g(y,\mu).
\label{eq:g-frozen}
\end{equation}
where $\mu\in\R^q$. The similarity of systems (\ref{eq:f-nonaut}) and (\ref{eq:g-nonaut-tau}) means similar methods can be used to understand how tipping in (\ref{eq:f-nonaut}) relates to tipping in (\ref{eq:g-nonaut-tau}). 

Assume that $\Lambda(s)$ is bi-asymptotically constant, i.e. there are constants
$$
\lambda_{\pm}:=\lim_{s\rightarrow \pm \infty} \Lambda(s).
$$
For $\lambda=\lambda_-$ the frozen {\em past limit system} for (\ref{eq:f-nonaut},\ref{eq:g-nonaut}) is
\begin{equation}
\left.
\begin{aligned}
    \frac{d}{dt} x & = f(x,\lambda)\\
    \epsilon\frac{d}{dt} y & = g(y,M(x))
\end{aligned}\right\}.
\label{eq:slowfastodefrozen}
\end{equation}
We suppose there is a linearly stable equilibrium $(X_-,Y_-)$, and we start within a neighbourhood of this equilibrium. By results in \cite{ashwin2017parameter} there is a unique trajectory 
$(\tilde{x}(t),\tilde{y}(t))$ that is a
pullback point attractor of the nonautonomous system (\ref{eq:f-nonaut},\ref{eq:g-nonaut}) with past limit
$$
\lim_{t\rightarrow -\infty}(\tilde{x}(t),\tilde{y}(t)))=(X_-,Y_-)
$$
and this will attract all initial conditions close enough to $(X_-,Y_-)$ in the far past (i.e. in pullback sense).
Note that $\tilde{x}$ depends on $\Lambda$ and $r$ while $\tilde{y}$ depends on $\Lambda$, $r$ and $\epsilon$.

The question of whether the system starting at $(X_-,Y_-)$ will undergo B-tipping is determined by the behaviour of this pullback attractor, and whether there is bifurcation-induced tipping of $x$ or $y$.
In a forthcoming paper \cite{ritchie2025cascades} we demonstrate how tipping for a pullback attractor of (\ref{eq:f-nonaut},\ref{eq:g-nonaut}) can be understood in terms of the dynamics of the frozen system (\ref{eq:slowfastodefrozen}) and how this is affected by the coupling. Note that we can view the full system as a pair of nonautonomous systems
(\ref{eq:f-nonaut},\ref{eq:g-nonaut}) and the frozen system (\ref{eq:slowfastodefrozen}) as a system (\ref{eq:f-frozen},\ref{eq:g-nonaut}).

\begin{figure}
    \centering
    \includegraphics[width=0.9\linewidth]{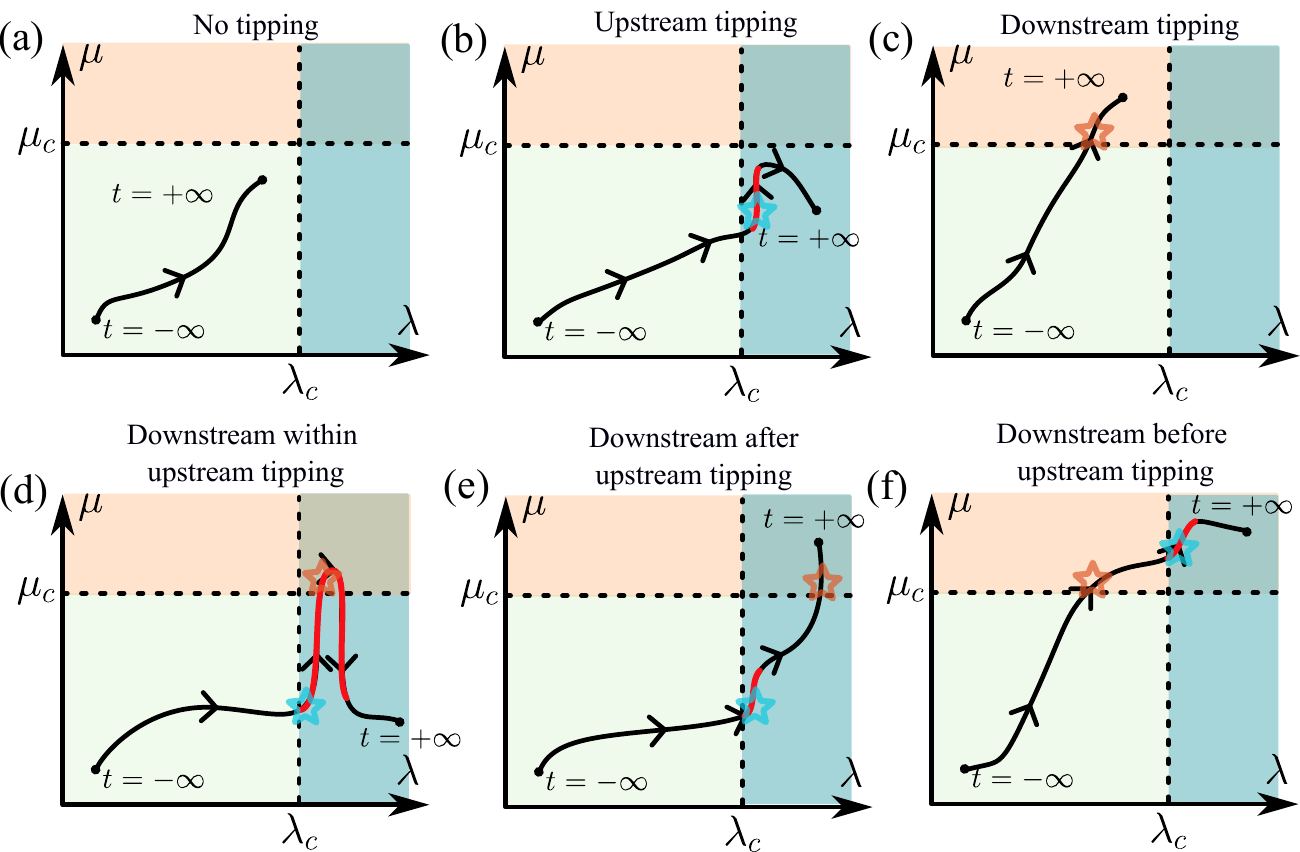}
    \caption{Schematic diagram showing possible paths (in black/red) of $(\Lambda(rt),M(\tilde{x}(t)))$ in the $(\lambda,\mu)$ plane for a pullback attractor of (\ref{eq:f-nonaut},\ref{eq:g-nonaut}). Crossing $\lambda=\lambda_c$ leads to upstream tipping (blue star), and crossing $\mu=\mu_c$ leads to downstream tipping (orange star). The light green region represents (in the limit $r$ and $\epsilon$ small) to a ``safe operating space'' where no tipping occurs, blue corresponds to upstream tipping, orange to downstream tipping and grey to both. The red part of the path indicates rapid motion in $\tilde{x}$ due to an upstream tipping. We show cases for slow variation of parameters and an accelerating cascade with (a) no tipping (b) upstream tipping only (c) downstream tipping only (d) downstream-within-upstream tipping (e) downstream after upstream tipping (f) downstream before upstream tipping. In case (d), extrapolation of an early warning to crossing $\mu=\mu_c$ will only be valid during tipping of the upstream system.}
    \label{fig:schem-path}
\end{figure}

In the limit of small $r$ and $\epsilon$, tipping behaviour can be studied in terms of the path $(\Lambda(rt),M(\tilde{x}(t))$ in the $(\lambda,\mu)$-plane as shown schematically in Figure~\ref{fig:schem-path}. Suppose there is a bifurcation at $\lambda=\lambda_c$ where the branch containing $X_-$ undergoes B-tipping (in the frozen upstream system~\eqref{eq:f-frozen}) and a bifurcation at $\mu=\mu_c$ where the branch containing $Y_-$ undergoes B-tipping (in the frozen downstream system~\eqref{eq:g-frozen}), then the tipping behaviour is determined by how the forcing path crosses $\lambda=\lambda_c$ or $\mu=\mu_c$. Figure~\ref{fig:schem-path} illustrates possible cases where upstream and downstream tipping may or may not occur, and highlights the sequence in which they occur. The blue star shows where upstream tipping occurs (which leads to a period of faster change of $M(\tilde{x}(t))$) while the orange star shows where downstream tipping occurs.
Note that there may be (a) no tipping (b) only upstream (c) only downstream tipping, or (e,f) where one tipping follows the other over timescales determined by the external forcing. If there is a sequence of tippings then we note this may or may not follow the direction of coupling.

The case (d) in Figure~\ref{fig:schem-path} that we denote {\em downstream-within-upstream tipping} is particularly interesting; in this case downstream tipping starts and finishes while upstream tipping is still underway. This presents a challenge to finding a skilful early warning prediction of downstream tipping in that extrapolation can only be expected to be accurate for a period of time while the upstream tipping event is underway. We note this feature of an accelerating cascade cannot be distinguished in the taxonomy of \cite{Klose2021Cascade} as it requires a timescale separation in the coupled elements.

\subsection{An example with coupled tipping elements}

We consider a system where upstream and downstream tipping elements are one-dimensional systems with hysteresis, to illustrate the discussion of the previous section. For $x\in\R$ and $y\in\R$ let
\begin{equation}
    \left.
    \begin{aligned}
        \frac{d}{dt} x=& f(x,\Lambda(rt))\\
        \epsilon\frac{d}{dt} y=&f(y,M(x))
    \end{aligned}
    \right\}
    \label{eq:model}
\end{equation}
where, as before, $f(x,\lambda) : = 3x-x^3+\lambda$ is bistable for $-2<\lambda<2$ and where, as before, there are fold bifurcations at $\lambda_u=2$, $\lambda_l=-2$. We assume $r>0$ and $\epsilon>0$ are small. We denote by $X_u$ and $X_l$ the upper and lower stable branches of equilibria for the frozen subsystem
$$
\dot{x}=f(x,\lambda)
$$
at $x=X_{l}(\lambda)$ for $\lambda<\lambda_u$ and $x=X_{u}(\lambda)$ for $\lambda>\lambda_l$. These are connected by an unstable branch $x=U(\lambda)$ for $\lambda_l<\lambda<\lambda_u$, where $X_l(\lambda)<U(\lambda)<X_u(\lambda)$. 
We force the system (\ref{eq:model}) via the monotonic 
parameter shift (\ref{eq:monotonic}) and consider a coupling function that includes linear and localised coupling:
\begin{equation}
M(x)=a_1+a_2x+a_3\bump(a_4(x-a_5)),
\label{eq:coupling}
\end{equation}
with $\bump(x)$ defined as in \eqref{eq:bump}. Depending on the values of $a_1,\ldots,a_5$, the behaviour of $M(x)$ can drive the downstream system over a tipping point. 
We write pullback attracting trajectory limiting to the lower branch $X_l(\lambda_-)$ in the limit $t\rightarrow-\infty$ as $x^{[r]}(t)$ 
and define 
$$
X_{-}:=\lim_{t\rightarrow -\infty}x^{[r]}(t) \mbox{ and }
X_{+}^{[r]}:=\lim_{t\rightarrow \infty}x^{[r]}(t).
$$
Note that $X_-=X_l(\lambda_-)$. Using the monotonic nature of $\Lambda$ we can classify the behaviour of $x^{[r]}(t)$ as follows:
\begin{itemize}
    \item  If $\lambda_+<\lambda_u$ then for all $r>0$ there will be end-point tracking; i.e., $X_+^{[r]}=X_l(\lambda_+)$.
    \item If $\lambda_+>\lambda_u$ then for all $r>0$ there will be B-tipping; i.e., $X_+^{[r]}=X_u(\lambda_+)$.
\end{itemize}
We write the effective forcing of the $y$ dynamics as
$$
\mu^{[r]}(t):=M(x^{[r]}(t))
$$
and define
$M_-:=M(X_-)=\lim_{t\rightarrow-\infty} \mu^{[r]}(t)$.
If we assume that $M_-<\lambda_u$ then for any $r>0$, $\epsilon>0$ there is a unique (pullback attracting) trajectory $(x^{[r]}(t),y^{[r,\epsilon]}(t))$ of (\ref{eq:model}) where $y^{[r,\epsilon]}(t)$ limits in the past to $X_l(M_-)$: namely
\begin{equation}
Y_-:=X_l(M_-)=\lim_{t\rightarrow -\infty} y^{[r,\epsilon]}(t).
\end{equation}
We can classify the behaviour of the downstream system $y^{[r,\epsilon]}(t)$ as tracking or tipping purely in terms of the effective forcing. Given $x^{[r]}(t)$ we define 
$$
M_+=\lim_{t\rightarrow \infty} \mu^{[r]}(t).
$$
Note that $M_+$ may depend on $r$.
If $\sup_{t} \mu^{[r]}(t)<\lambda_u$ then we have end-point tracking; i.e., $\lim_{t \rightarrow \infty} y^{[r,\epsilon]}(t)=X_l(M_+)$ for small enough $\epsilon$.  If $M_+>\lambda_u$ then we have B-tipping; i.e., $\lim_{t \rightarrow \infty} y^{[r,\epsilon]}(t) = X_u(M_+)$ for small enough $\epsilon$. If $M_+<\lambda_u$ and there is a $t$ (depending on $r$) such that $\mu^{[r]}(t)>\lambda_u$ then for all $r$ and small enough $\epsilon$ there is B-tipping for the downstream system, while for large enough $\epsilon$ there is end-point tracking.
Observe that threshold overshoots can occur if $\mu^{[r]}(t)$ is non-monotonic. This is the case where for large enough $\epsilon$ one can have a safe overshoot as discussed in \cite{ritchie2019inverse}. For a more detailed discussion of the dynamics of this system in terms of its bifurcation structure, we refer to \cite{ritchie2025cascades}.

\subsection{EWS for prediction of downstream tipping}
\label{sec:EWS}

Let us now consider a stochastic version of (\ref{eq:model}), namely: 
\begin{equation}
\begin{aligned}
dx &= f(x,\Lambda(rt))dt + \eta_1 dV_t \\
\epsilon\, dy & = f(y,M(x))dt+\eta_2 dW_t
\end{aligned}
\label{eq:modelsde}
\end{equation}
with $\eta_i$ small. We consider $r$ and $\epsilon$ small quantities representing timescale separation, $V_t$ and $W_t$ are independent standard Wiener processes with amplitudes $\eta_i$ (we consider only additive isotropic noise for simplicity here).  Consider forcing as in (\ref{eq:monotonic}) and coupling as in (\ref{eq:coupling}) and choose default parameters
\begin{equation}
\begin{array}{c}
\lambda_-=0,~\lambda_+=4,\\
a_1=a_2=0,~a_3=2.5,~a_4=2,~a_5=0.5,\\
\epsilon=0.05,~r=0.05,~ \eta_1=\eta_2=0.1
\end{array}
\label{eq:defaultLambdaM}
\end{equation}
that determine the forcing of the upstream system and the coupling from upstream to downstream systems.
We compute early warning predictors for upstream and downstream systems as outlined in Section~\ref{sec:class}. Early warning indicators are calculated over some time interval up to $t_p$ and the linear fit to $\alpha^2(t)$ over a fitting range of $t_p$ is extrapolated to time $t_p+\Delta$: this value is used as a scoring classifier for tipping in the interval from $t_p$ to $t_p+\Delta$, and ROC/AUC calculated as discussed in Section~\ref{sec:EWSskill}.

Recall that Figure~\ref{fig:roc_auc_upstream} shows ROC/AUC curves for the upstream system where $\lambda_+$ is assumed to be chosen randomly with equal probability from $\{1,3\}$,
such that roughly half realisations will avoid tipping while the other half tip. That figure suggests that the predictor provides a skilful prediction for detecting tipping in the upstream system for short enough prediction windows. The optimal threshold approaches zero for prediction times $t_p$ close to the crossing of the bifurcation point and for moderate time horizons $\Delta$. 

\begin{figure}
    \centering
    \includegraphics[width=\textwidth]{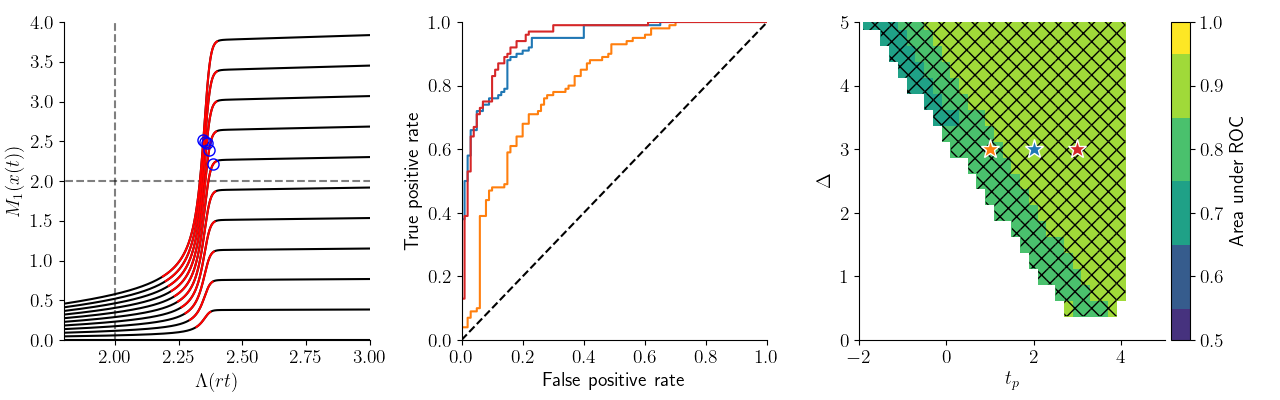}
    \caption{Quantifying the skill of prediction at time $t_p$ for downstream subsystem, $y$, of (\ref{eq:modelsde},\ref{eq:monotonic},\ref{eq:coupling}) for linear coupling. We take parameters \eqref{eq:defaultLambdaM} except $a_3=0$, $a_2$ is chosen from $\mathcal{U}[0,1]$ (and $a_1=a_2\sqrt{3})$) for each Monte-Carlo simulation. Left panel; shows $\Lambda(rt)$ against $M(x(t))$ shown red if  $|\frac{d}{dt} M(x(t))|>10^{-3}$. Blue circles indicate the moment of downstream system tipping, defined as when $y=0$. Grey dashed lines indicate thresholds at $\Lambda(rt) = 2$ and $M(x(t)) = 2$. Middle/right: We forecast tipping within the window $[t_p,t_p+\Delta]$ using the early warning predictor.
    Middle panel shows ROC curves for prediction times $t_p$, and time horizons $\Delta$ indicated by the coloured stars in the right panel. Right: Colour plot of area under ROC curve (AUC) for different prediction times, $t_p$ and time horizons $\Delta$. To construct the curves, 1,000 
    simulations were performed (with $a_1$ and $a_2$ randomly sampled). Then, for each prediction time $t_p$ and horizon $\Delta$, out of those 1,000 simulations an ensemble was created by picking the first 100 that show tipping and the first 100 that do not show tipping within the time window $[t_p, t_p+\Delta]$. If such an ensemble cannot be created, no ROC/AUC analysis was 
    performed (white areas in right panels). The ROC is calculated using these simulations and systematically considering a sufficiently wide range of thresholds, $\kappa$, such that for the 
    maximum $\kappa$ all trajectories are predicted to tip and the minimum $\kappa$ all trajectories are predicted not to tip. 
    Hatching denotes the value of the optimal threshold, $\kappa^{Opt}(t_p,\Delta)$: Cross hatching,  $|\kappa^{Opt}(t_p,\Delta)|>10$; star hatching, $5\leq |\kappa^{Opt}(t_p,\Delta)|<10$; dot hatching, $2\leq |\kappa^{Opt}(t_p,\Delta)|<5$; no hatching, $|\kappa^{Opt}(t_p,\Delta)|<2$.
    }
    \label{fig:roc_auc_downstream_lin}
\end{figure}

Figure~\ref{fig:roc_auc_downstream_lin} shows (left) downstream vs upstream forcing as well as (middle) ROC and (right) AUC for the downstream system with linear coupling and a scenario with first upstream tipping and then downstream tipping (scenario (e) in Figure~\ref{fig:schem-path}), i.e. parameters are as in (\ref{eq:defaultLambdaM}) except that $a_3=0$, $a_2 \sim \mathcal{U}[0,1]$ (and $a_1 = a_2\sqrt{3}$ for each Monte-Carlo simulation). This is chosen such that there is roughly an even split between the number of realisations that undergo tipping and those that do not. Interestingly, the early warning predictor applied to the downstream system still provides some skilful prediction, although degraded relative to the upstream system. This is due to the large variation in peak amplitudes. This is further reflected by the optimal scoring classifier being far away from zero. 

\begin{figure}
    \centering
    \includegraphics[width=\textwidth]{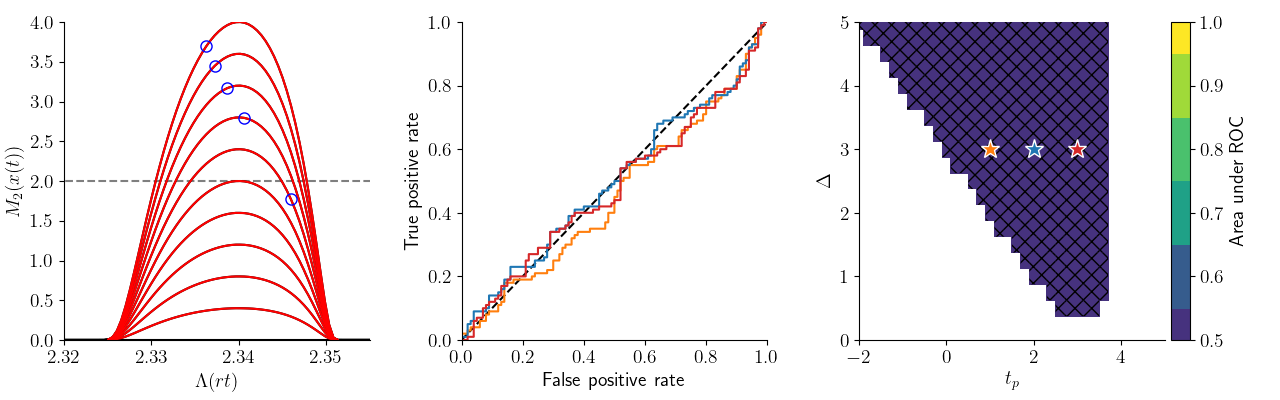}
    \caption{Quantifying the skill of prediction at time $t_p$ for downstream subsystem, $y$ as in Figure~\ref{fig:roc_auc_downstream_lin} but with localised coupling. In this case we consider default parameters (\ref{eq:defaultLambdaM}) except $a_3$ is drawn from $\mathcal{U}[0,4]$ for each Monte-Carlo simulation. Left panel shows $\Lambda(rt)$ against $M(x(t))$ shown red if  $|\frac{d}{dt} M(x(t))|>10^{-3}$. Blue circles indicate the moment of downstream system tipping, defined as when $y=0$. Grey dashed line indicates the threshold at $M(x(t)) = 2$. 
    Middle panel shows ROC curves for prediction times $t_p$, and time horizons $\Delta$ indicated by the coloured stars in the right panel. 
    Right panel shows a colour plot of AUC for different prediction times, $t_p$ and time horizons $\Delta$.  To construct the curves, 1,000 simulations were performed (with $a_3$ randomly sampled). Then, for each prediction time $t_p$ and horizon $\Delta$, out of those 1,000 simulations an ensemble was created by picking the first 100 that show tipping and the first 100 that do not show tipping within the time window $[t_p, t_p+\Delta]$. If such ensemble cannot be created, no ROC/AUC analysis was performed (white areas in right panels).
    Hatching denotes the optimal threshold, $\kappa^{Opt}(t_p,\Delta)$: Cross hatching,  $|\kappa^{Opt}(t_p,\Delta)|>10$; star hatching, $5\leq |\kappa^{Opt}(t_p,\Delta)|<10$; dot hatching, $2\leq |\kappa^{Opt}(t_p,\Delta)|<5$; no hatching, $|\kappa^{Opt}(t_p,\Delta)|<2$. Default parameter values are otherwise used.
    }
    \label{fig:roc_auc_downstream_loc_bump}
\end{figure}

By contrast, Figure~\ref{fig:roc_auc_downstream_loc_bump} shows that all skill is lost if we consider localised coupling so that there is a change in $M(x(t))$ only during the tipping of $x(t)$ (i.e., scenario (d) in Figure~\ref{fig:schem-path}). In particular, we consider parameters as in (\ref{eq:defaultLambdaM})  except for $a_3 \sim \mathcal{U}[0,4]$. 
In this case the early warning predictor cannot differentiate between forcing that leads to tipping or not except during the tipping of the upstream system. Although the mechanism is still a fold bifurcation, the skill is no better than a completely random prediction. 

\section{Discussion}
\label{sec:discuss}

In the past, several authors (notably \cite{BenYami2023AMOCDataCSD}) have highlighted the range of factors that need to be considered for an early warning predictor of tipping points to be skilful. In this paper, we focused on the problem of extrapolation. Fixing the time horizon for early warnings makes the test more amenable to quantitative analysis, especially if there is a way to guarantee extrapolation over a short time horizon. It seems to be an important open problem to find ways to estimate the time of reliable extrapolation, for example, by bounded higher derivatives of any trend.

The methods we used for estimating stability in Section~\ref{sec:estimation} can certainly be improved, for example, using the maximum likelihood methods of \cite{Ditlevsen2023AMOCPrediction} for linear passage through a normal form rather than the linearised version, or semiparametric methods such as those proposed in \cite{kwasniok2024semiparametric}. Ideally, one can aim to estimate the normal form together with a model for the temporal variation of the forcing that will imply a fitting to a trend. However, the problem of extrapolating the trend will remain no matter how precisely the estimation and fitting problem are done.

Some takeaway messages of this study are the following:
\begin{itemize}
\item  {\em Early warnings are particularly susceptible to errors in extrapolation.}
\item {\em Lack of extrapolation may come from a variety of sources:}
\begin{itemize}
    \item {\em Parameter variation being nonlinear in time} (see Section~\ref{sec:examplenonmon}).
    \item {\em Being outside the region of validity of a normal form}
    (see Section~\ref{sec:examplevalidity}).
    \item {\em Presence of other eigenvalues} (see Section~\ref{sec:exampleeigenvalue}).
    \item {\em Multiple timescale behaviour of coupling from an upstream system undergoing tipping} (see Section~\ref{sec:propagation}).
\end{itemize}
\item {\em Specifying a time horizon makes the skill more easily quantifiable.}
\end{itemize}
In particular, a distribution of extrapolated times of tipping will only translate into a distribution of possible times of tipping if reliable extrapolation of the stability of the attracting state is possible.

We emphasise that, thus far, we have worked within a limited setting to highlight important factors that may spoil extrapolation. Namely, we considered only nonlinear and nonautonomous dynamical systems that are well represented by weak noise and slow drift through a fold bifurcation. 
It will be of great relevance to identify if there are additional factors for skilful warning of approach to other bifurcations, such as Hopf or transcritical. Also, non-smooth systems can be expected to have unreliable extrapolation, causing problems with early warnings \cite{budd2024dynamic}. We considered cases where tipping propagates from a slow upstream system to a faster downstream system under the assumption of timescale separation; it will be important to generalise this to cases where there is bidirectional coupling and/or more than two systems involved.

The insights in this paper are relevant for various systems -- and indicate that early warning prediction might be unfeasible or possible only very late for these systems.
First, climate forcing happens on multiple time scales \cite{von2021quantification} and forcing is only ever monotonic over some limited timescale.
Second, many systems seem to have multiple tipping points instead of tipping fully in one tipping event. Spatially heterogeneous systems in particular easily can have multiple tipping points via fragmented tipping \cite{bastiaansen2022fragmented}. For example, the melting of the Pine Island Glacier might be organised via three tipping points \cite{rosier2021tipping}.
Third, climate (sub)systems respond to forcing on many different time scales \cite{von2021quantification}. Hence, a stable state has many eigenvalues associated with it. It can happen that tipping is organised through the crossing of one eigenvalue that is typically not dominant. This happens easily in fast-slow systems, such as dryland ecosystems \cite{bastiaansen2020effect} or in energy balance models \cite{bastiaansen2023climate}.
Finally, various examples have been explored in the climate literature where a slow system can tip and influence a faster system with potential to tip (for an overview of tipping element interactions and their timescales, see e.g. \cite{wunderling2024climate}). For example, melting of Greenland ice mass can cause tipping of the AMOC through freshwater coupling (as explored using a conceptual model in \cite{sinet2024amoc}). This example has similar localised coupling as explored in this paper in the sense that when a tipping of the land ice is underway, this is when the freshwater flux into the North Atlantic is at a maximum.
All of these, and other, systems have properties that we have identified in this paper that can lead to the breakdown of early warning predictions due to lack of extrapolation -- and hence care should be taken when trying to predict their future.

There has been work on tipping cascades in other contexts; for example, \cite{ashwin2017fast} consider noise-induced tipping for coupled multistable systems with weak noise and show that there are various regimes of transmission of tipping. Such effects will also become important in cases where noise is large relative to changes in parameters. Moreover, for noise-induced tipping, precursors need to be associated with extreme behaviour of the noise rather than a slowly changing forcing crossing a threshold.

For a prediction of future tipping to be useful, it is often necessary for it to be neither too early nor too late. If there is a hope of adaptation to an impending tipping event, then some time is needed to prepare, and this will indicate when the warning is too late. If the warning is too early, then the cost of early adaptation may outweigh the benefits. However, as demonstrated in this study, the time window in which skilful early warning prediction can be made depends on the specifics of the system and forcing. In fact, we have described some settings in which early warning prediction seem unfeasible, such as the setting of tipping within tipping where a downstream tips during the tipping of an upstream system.

\subsection*{Acknowledgements}
We acknowledge support from the European Union's Horizon Europe research and innovation programme under grant agreement No. 101137601 (ClimTip): Funded by the European Union. We acknowledge support by the UK Advanced Research and Invention Agency (ARIA) via the project ``AdvanTip". A.S.vdH acknowledges funding by the Dutch Research Council (NWO) through the NWO-Vici project ``Interacting climate tipping elements: When does tipping cause tipping?'' (project VI.C.202.081). Views and opinions expressed are however those of the author(s) only and do not necessarily reflect those of the European Union or the European Climate, Infrastructure and Environment Executive Agency (CINEA) or other funders. Neither the European Union nor the granting authority can be held responsible for them. 
For the purpose of open access, the authors have applied a Creative Commons Attribution (CC BY) licence to any Author Accepted Manuscript version arising from this submission. 

\subsection*{Data accessibility statement}
The code to generate the data for the figures presented is available on
\url{https://github.com/peterashwin/acceleratingcascades}.

\bibliographystyle{plain}


\end{document}